\documentstyle[12pt,epsf]{article} 

\begin{document}

\title{Tunneling and Nucleation Rate in the 
$\left(\frac{\lambda}{4!}\phi^4+\frac{\sigma}{6!}\phi^6\right)_{3}$ Model} 

\author{Gabriel H. Flores$^{1}\;$\thanks{E-mail:
gflores@lafex.cbpf.br}$\;$,
Rudnei O. Ramos$^{2}\;$\thanks{E-mail: rudnei@symbcomp.uerj.br}$\;$ and
N. F. Svaiter$^{1}\;$\thanks{E-mail: nfuxsvai@lafex.cbpf.br}\\ \\
{\it $^1\;$ Centro Brasileiro de Pesquisas Fisicas-CBPF,}\\
{\it Rua Dr.Xavier Sigaud 150, 22290-180 Rio de Janeiro, RJ, Brazil} 
\\
{\it $^2 \;$ Universidade do Estado do Rio de Janeiro,}\\
{\it Instituto de F\'{\i}sica - Departamento de F\'{\i}sica Te\'orica,}\\
{\it 20550-013 Rio de Janeiro, RJ, Brazil}} 

\date{February 1999} 

\maketitle

\begin{abstract} 

We evaluate both the vacuum decay rate at zero temperature
and the finite temperature nucleation rate for the 
$\left(\frac{\lambda}{4!}\phi^4+\frac{\sigma}{6!}\phi^6\right)_{3D}$ model.
Using the thin-wall approximation, we obtain the bounce solution for the 
model and we were also able to give the approximate eigenvalue equations
for the bounce. 

\vspace{0.34cm} 
\noindent 
PACS number(s):  11.27.+d,11.90.+t  

\end{abstract} 

\newpage

\section{Introduction}

The scalar field model with potential $U(\phi)=\frac{m^2}{2}
\phi^2 + \frac{\lambda}{4 !} \phi^4 + \frac{\sigma}{6 !} \phi^6$ is the
simplest model exhibiting a rich phase structure and
for studying tricritical phenomena in both two and
three dimensional systems [see, {\it e.g.} Refs. \cite{amaral} and 
\cite{novo}]. 
In terms of the phase diagram for the
model, a tricritical point can emerge whenever we have three phases
coexisting simultaneously. {}For the above potential, in the absence of
corrections due to fluctuations, one can have a second order transition
in $m$ when the scalar field mass vanishes and $\lambda >0, \sigma >0$.
A first order transition happens for the case of $\lambda < 0, \sigma >
0$. The tricritical point occurs when the quartic coupling constant
vanishes (with $m=0, \sigma>0$). A study of the phase diagram for this
model in $3D$, at finite temperature, was recently done \cite{nami} and it
has been shown that there is a temperature $\beta^{-1}
(m,\lambda,\sigma)$ for which the physical thermal mass $m_\beta$ and
coupling constant $\lambda_\beta$ vanishes, thus characterizing the
tricritical point.

There are many possible applications associated with the model with
potential $U(\phi)$ with $\phi^6$ interaction. {}For instance, in $D=2$,
it is known \cite{zamolo} that the minimal conformal quantum field
theory, with central charge $7/10$ (the tricritical Ising model) is in
the same universality class, in the scaling region near the tricritical
point, of the Landau-Ginzburg model with the above potential. This
$\phi^6$ potential model can then be thought of as the continuum
realization of the Ising model with possible applications in, {\it
e.g.}, the description of adsorbed helium on krypton-plated graphite
\cite{lebow}, in understanding the statistical mechanics of binary
mixtures, such as $He^3 \, - \, He^4$ \cite{griff}, etc. These are just
a few examples of systems exhibiting tricritical phenomena in condensed
matter physics.

In field theory in general, the $\phi^6$ model has been used in the study 
of polarons, or solitonic like field configurations on systems of low dimensionality
\cite{amaral}. Also, a gauged ($SU(2)$) version of the $\phi^6$ potential model in Euclidean
$3D$, has recently been used to study a possible existence of a
tricritical point in Higgs models at high temperatures \cite{arnold}. In
this context the tricritical point is characterized by the ratio of the
quartic coupling constant and the gauge coupling constant of the
effective three-dimensional theory, obtained from the $3+1\, D$ high
temperature, dimensionally reduced $SU(2)$ Higgs model. This can be
particularly useful in the context of the study of the electroweak phase
transition.

There are then many reasons that make the $\phi^6$ potential model an
interesting model to be studied. In this paper, we will be particularly
interested in studying the regime of parameters for which:

\begin{equation}
m^2 > 0\,, \;\lambda < 0\,,\;\sigma > 0 \;\; {\rm and} \;\;
\left[\left(\frac{\lambda}{3!}\right)^2-
4 m^2 \, \frac{\sigma}{5!}\right] > 0 \;.
\label{param}
\end{equation}

\noindent
In this case $U(\phi)$ has three relative
minima $\phi_{t\pm}$ and $\phi_f$ (see Fig. 1). {}For these parameters
the system has metastable vacuum states and it may exhibit a first order
phase transition. The states of the classical field theory for which
$\phi=\phi_{t\pm}$ are the unique classical states of lowest energy
(true vacuum) and, at least in perturbation theory, they correspond to
the unique vacuum states of the quantum theory. The state of the
classical field theory for which $\phi=\phi_f$ is a stable classical
equilibrium state. However, it is rendered unstable by quantum effects,
{\it i.e.}, barrier penetration (or over the barrier thermal
fluctuations, at finite temperatures). $\phi_f$ is the false vacuum
(the metastable state).

We will compute the vacuum decay (tunneling) rate at both zero and
finite temperature. {}For calculation reasons we will restrict ourselves
to the thin wall approximation for the true vacuum bubble (or bounce
solution). We then consider the energy-density difference between the
true and false vacuum as very small as compared with the height of the
barrier of the $U(\phi)$ potential \cite{coleman,linde}. {}From this, we
are able to give the explicit expression for the bounce and also
to qualitatively describe the eigensolutions for the bounce 
configuration. The paper is
organized as follows: In Sec. II we obtain the bounce field configuration
for the
model and we compute the Euclidean action in the thin wall approximation.
In Sec. III we calculate the vacuum decay rate at zero temperature and  we
discuss the eigenvalue equations obtained for the bounce configuration within 
the approximations we have taken for the bounce.
In Sec. IV we compute the nucleation rate at
finite temperature, following the procedure given in \cite{GMR}. In Sec.
V the concluding remarks are given. In this paper we use $\hbar=c=k_{b}=1$.

\section{The Vacuum Decay Rate and the Bounce Solution}

Let us consider a scalar field model, in
three-dimensional space-time, with Euclidean action given by

\begin{equation}
S_E(\phi)=\int d^3 x_E \left[ \frac{1}{2}\left(\partial_{\mu}\phi\right)^2 +
U(\phi)\right]\;,
\label{action}
\end{equation}

\noindent 
with 

\begin{equation}
U(\phi)=
\frac{m^2}{2}\phi^2+\frac{\lambda}{4!}\phi^4+\frac{\sigma}{6!}\phi^6 \;.
\label{pot}
\end{equation}

\noindent
As discussed in the introduction, we are interested in the regime for
which the parameters in (\ref{pot}) satisfies (\ref{param}),
such that the potential $U(\phi)$ exhibits non-degenerate local minima and
then metastable states. The picture we have in mind is that once the system is
prepared in the false vacuum state, it will evolve to the true vacuum state
by tunneling (at zero temperature) or by bubble nucleation, triggered by
thermal fluctuations over the potential barrier.

In the case of quantum field theory at zero temperature the study of the
decay of false vacuum was initiated by Voloshin, Kobsarev and Okun
\cite{Voloshin} and later by Callan and
Coleman \cite{coleman,callan}, who developed the so called bounce method
for the theory of quantum decay. In this context the decay rate per unit
space-time volume $V_3$ is given by

\begin{equation} 
\frac{\Gamma}{V_3}=\left(
\frac{\Delta S_{E}}{2\pi}\right)^{3/2}
\left[\frac{{\rm det}'\left(-\Box_{E}+U''(\phi_b)\right
)}{{\rm det}\left(-\Box_{E}+U''(\phi_f)\right)}\right]^{-1/2}
e^{-\Delta S_{E}}\left(1+{\cal O}(\hbar)\right) \;,
\label{decayrate} 
\end{equation} 

\noindent 
where  $\Box_{E}=
\frac{\partial^2}{\partial{\tau^2}}+\frac{\partial^2}{\partial{x^2}}
+\frac{\partial^2}{\partial{y^2}}$,
$U''(\phi)=\frac{d^2U(\phi)}{d\phi^2}$ and  
$\Delta S_E = S_E (\phi_b) -S_E(\phi_f)$.
$S_{E}(\phi_b)$ is the
Euclidean action evaluated at its extreme
(specifically a saddle point), $\phi=\phi_b$, where $\phi_b$ is the
bounce: a solution of the field equation of motion, $\delta S_E/\delta \phi
|_{\phi = \phi_b} =0$, with
the appropriate boundary conditions.  
The prime in the determinantal prefactor in (\ref{decayrate}) means that the
three zero eigenvalues (the translational modes) 
of the $[-\Box_{E}+U''(\phi_b)]$ operator has been
removed from it.

As it was shown by Coleman, Glaser and
Martin \cite{martin}, the solution that minimizes $S_E$ is a spherical
symmetric solution, $r^2=\tau^2+x^2+y^2$ (in $D$ dimensions the solution
has $O(D)$ symmetry) and then $\phi_b$ can be written as the solution of
the radial equation of motion 

\begin{equation}
\frac{d^2\phi_b}{dr^2}+ \frac{2}{r} \frac{d\phi_b}{dr}=U'(\phi_b) \;,
\label{equation}\label{b}
\end{equation} 

\noindent
with the boundary conditions: $\lim_{r\rightarrow
\infty}\phi_b(r)=\phi_f$ and $\frac{d\phi_b}{dr}|_{r=0}=0$.
 
At finite temperature the calculation of the decay rate were first
considered in \cite{affleck} in the context of quantum mechanics and
later by Linde \cite{linde}, for quantum field theory. In \cite{linde},
it is argued that temperature corrections to the nucleation rate are
obtained recalling that finite temperature field theory (at sufficiently
high temperatures) in $D=d+1$ dimensions is equivalent to d-dimensional
Euclidean quantum field theory with $\hbar$ substituted by $T$. 
At finite temperature, the bounce, $\phi_B \equiv \phi_B(\rho)$ ($\rho = 
|{\bf r}|$), is a static solution of the field equation
of motion:

\begin{equation}
\frac{d^2\phi_B}{d\rho^2}+ \frac{1}{\rho} \frac{d\phi_B}{d\rho}=
U'(\phi_B)\; ,
\label{B}
\end{equation}

\noindent 
with boundary conditions: $\lim_{\rho\rightarrow
\infty}\phi_B(\rho)=\phi_f$ and $\frac{d\phi_B}{d\rho}|_{\rho=0}=0$. 
The vacuum decay rate, or bubble nucleation rate in this case, is 
proportional to $e^{-\Delta E /T}$, where $\Delta E$ is the nucleation
barrier, given by ($\beta = 1/T$)

\begin{equation}
\frac{\Delta E}{T} = \beta \int d^2 {\bf x} \left[{\cal L}_E (\phi_B) -
{\cal L}_E (\phi_f) \right] \;,
\label{deltaE}
\end{equation}

\noindent
where ${\cal L}_E$ is the Euclidean Lagrangian density. 

The problem of the computation of the nucleation rate 
at finite temperature was recently 
reconsidered by
Gleiser, Marques and Ramos \cite{GMR}, who have used the early works of
Langer \cite{langer}. In this context the nucleation rate is given by

\begin{equation}
\left(\Gamma\right)_{\beta}=-\frac{|E_-|}{\pi}
{\rm Im}\left[\frac{{\rm det} \left[-\Box_{E}+U''(\phi_B)\right
]_{\beta}}{{\rm det} \left[-\Box_{E}+U''(\phi_f)
\right]_{\beta}}\right]^{-1/2}
e^{-\Delta E/T}\left(1+{\cal O}(\hbar)\right) \;,
\label{reconcha}
\end{equation}

\noindent
where $E_-^2$ is the negative eigenvalue associated to the 
operator $[-\Box_{E}+U''(\phi_B)]$.

To compute the vacuum decay rates, we need, therefore, to solve Eq.
(\ref{b}) for $\phi_b(r)$ (at zero temperature), or Eq. (\ref{B}) for 
$\phi_B(\rho)$ (at finite temperature). In the
model studied here, we can give an approximate analytical treatment for
the bounce solution in the so called thin-wall approximation, in which case
the energy-density difference between the false and true vacuum states can be
considered very small, as compared to the height of the potential
barrier: $\epsilon_0=U(\phi_f)-U(\phi_t)\ll U(\phi_2)$ (see Fig. 1). By
interpreting $\phi_b$ as a position and $r$ as the time, Eq. (\ref{b})
can be seen as a classical equation of motion for a particle moving in a
potential $-U(\phi)$ and subject to a viscous like damping force, with
stokes's law coefficient inversely proportional to the time. The 
particle motion can then be interpreted as it was released at rest,
at time zero
(because of the boundary condition $\frac{d\phi_b}{dr}|_{r=0}=0$).
{}From (\ref{b}) it follows that

\begin{equation}
\frac{d}{dr}\left[\frac{1}{2}\left(\frac{d\phi_b}{dr}\right)^2
-U(\phi_b)\right]=-\frac{2}{r}\left(\frac{d\phi_b}{dr}\right)^2
\leq 0 \;,
\label{duck}
\end{equation}

\noindent
meaning that the particle loses energy. Then, if the initial position of 
the particle is chosen to be at the left of some value $\phi=\phi_1$ 
(see Fig. 2), it will never reach the position $\phi_f$. 
Now, for $\phi_b(r)$ very close to $\phi_t$, we can  linearize 
Eq. (\ref{b}) to obtain

\begin{equation}
\left[\frac{d^2}{dr^2}+\frac{2}{r}\frac{d}{dr}-\mu^2 \right]
(\phi_b-\phi_t)=0 \;,
\label{lineal}
\end{equation}

\noindent
where $\mu^2=U(\phi_t)$. The solution of (\ref{lineal}) is 

\begin{equation}
\phi_b-\phi_t=(\phi(0)-\phi_t)\frac{\sinh(\mu r)}{\mu r}\;.
\end{equation}

\noindent
Therefore, if we choose the position of the particle to be initially 
sufficiently
close to $\phi_t$, we can arrange for it to stay arbitrarily close 
to $\phi_t$ for arbitrarily large $r$. 
But for sufficiently large $r$, $r=R$, the viscous damping force 
can be neglected, since its coefficient is inversely proportional 
to $r$. And if the viscous damping is neglected, the particle will 
reach the position $\phi_f$ at a finite time $R+\Delta R$. Then,
by continuity, there must be an initial position between $\phi_t$ and 
$\phi_1$ for which the particle will come to rest at $\phi_f$, 
after an infinity time.
{}From the above arguments, we can find a general expression for $\phi_b$,
valid for $\epsilon_0\ll U(\phi_2)$. 
In order to not lose too much energy, we must choose $\phi_b(0)$, 
the initial position of the particle, very close to $\phi_t$. 
The particle then stays close to $\phi_t$ until some very large time, 
$r=R$. Near $R$ (between $R-\Delta R$ and $R+\Delta R$, 
$\Delta R\ll R$) 
the particle moves quickly (according to Eq. (\ref{b}), 
neglecting the viscous damping force), through the valley in Fig. 2 and 
it slowly comes to rest at $\phi_f$, after an infinity time. 
Thus, we can write for the bounce the following expression ($\phi_f=0$) 

\begin{equation}
\phi_b(r)=\left\{
\begin{array}{ll} 
\phi_t & \mbox{$0<r<R-\Delta R$} \\
\phi_{\rm wall}(r-R) & \mbox{$R-\Delta R<r<R+\Delta R$} \\
\phi_f & \mbox{$R+\Delta R<r<\infty$}
\end{array} 
\right. \;,
\label{bounce}
\end{equation}

\noindent
where $\phi_{\rm wall}$ satisfies the equation 

\begin{equation}
\frac{d^2\phi_{\rm wall}}{dr^2}=U'(\phi_{\rm wall})
\label{w}\;.
\end{equation}

\noindent
Eq. (\ref{bounce}) is the thin-wall approximation for the 
bounce solution. {}From Eq. (\ref{w}), we obtain

\begin{equation}
\int_{0}^{\phi_{\rm wall}} \frac{d\phi}{\sqrt{2U(\phi)}}=r \;.
\label{pato}
\end{equation}

\noindent
By rewriting $U(\phi)$ as

\begin{equation}
U(\phi)=\frac{\sigma}{6!}\phi^2{\left ( \phi^2-\phi_0^2\right)}^2-\gamma
\frac{\phi^2}{\phi_0^2}\;, 
\label{pinga}
\end{equation}

\noindent
with

\begin{equation}
\phi_0^2=-\frac{1}{2}\frac{6!}{\sigma}\frac{\lambda}{4!}
\end{equation}
\label{algo}

\noindent
and

\begin{equation}
\gamma=\frac{\phi_0^2}{4}\left[\frac{6!}{\sigma}
{\left(\frac{\lambda}{4!}\right)}^2-2m^2\right] \;,
\label{delta}
\end{equation}

\noindent
then, by neglecting in  (\ref{pinga}) 
the term proportional to $\gamma$ 
(valid in the thin-wall approximation), we obtain that

\begin{equation}
\int_{0}^{\phi_{\rm wall}} \frac{d\phi}{\sqrt{2\frac{\sigma}{6!}}
\phi\left(\phi^2-\phi_0^2\right)}=r \;.
\label{pato2}
\end{equation}

\noindent
The above integral is straightforward and the solution for 
$\phi_{\rm wall}(r)$ can be written as

\begin{equation}
\phi^2_{\rm wall}(r)=\frac{\phi_0^2}{1+\exp{\left(\sqrt{8\frac{\sigma}
{6!}}\phi_0^2r\right)}} \;.
\label{wall}
\end{equation}

\noindent
This solution is shown in Fig. 3. 

Using (\ref{wall}), we obtain for $\Delta S_{E}$ the expression 

\begin{eqnarray}
\Delta S_{E} &=& 4\pi\int_{0}^{\infty}drr^2\left[\frac{1}{2}
\left(\frac{d\phi_b}
{dr}\right)^2+U(\phi_b)\right] \nonumber \\
&=& 4\pi\int_{0}^{R-\Delta R}drr^2U(\phi_t)+4\pi
\int_{R+\Delta R}^{\infty}drr^2U(0)+\nonumber \\ 
& & 4\pi\int_{R-\Delta R}^{R+\Delta R}drr^2 
\left[\frac{1}{2}{\left(\frac{d\phi_{\rm wall}}{dr}\right)}^2+
U(\phi_{\rm wall})\right] \;.
\label{porra}
\end{eqnarray}

\noindent
Since $ \Delta R \ll R$, in the last
integral of (\ref{porra}) we can take $r\approx R$ and we obtain

\begin{equation}
\Delta S_{E} \simeq -\frac{4}{3}\pi\epsilon_0 R^3+4\pi R^2 {\cal S}_0 \;,
\label{acao}
\end{equation}

\noindent
with ${\cal S}_0$, the bounce surface energy density, given by

\begin{equation}
{\cal S}_0=\int_{\phi_t}^{\phi_f}d\phi\sqrt{2U(\phi)}\approx
\int_{\phi_0}^{0}d\phi\sqrt{2\frac{\sigma}{6!}}\phi
\left(\phi^2-\phi_0^2\right)\;,
\label{S1}
\end{equation}

\noindent
where we have neglected in $U(\phi)$ the term $\gamma\phi^2/\phi_0^2$ 
and we also used $\phi_t\approx\phi_0$. 
Evaluating the above integral, we obtain for $S_0$ the result:

\begin{equation}
{\cal S}_0 \simeq \frac{\phi_0^4}{4}\sqrt{2\frac{\sigma}{6!}}\; .
\label{valor}
\end{equation}

In the next two sections, we deal with the evaluation of the 
determinantal prefactor appearing in Eqs. (\ref{decayrate}) and 
(\ref{reconcha}) and we obtain the subsequent radiative (1-loop) corrections
to (\ref{acao}).

\section{The Vacuum Decay Rate at $T=0$}

Let us consider, initially, the eigenvalue equations for the 
differential operators
appearing in (\ref{decayrate}):

\begin{equation}
\left[-\Box_{E} +U''(\phi_b)\right]\psi_b(i)=E_b^2(i)\psi_b(i)
\label{papi}
\end{equation}

\noindent
and 

\begin{equation}
\left[-\Box_{E}
+U''(\phi_f)\right]\psi_f(j)=E_f^2(j)\psi_f(j) \;.
\label{mami}
\end{equation}

\noindent
We then have for the determinantal prefactor of (\ref{decayrate}), the 
following
 
\begin{eqnarray}
K&=&\left[\frac{{\rm det}'\left[-\Box_E+U''(\phi_b)\right
]}{{\rm det}\left[-\Box_E+U''(\phi_f)\right]}\right]^{-1/2}
\nonumber\\
&=&\exp\left\{-\frac{1}{2}\ln\left[\frac{{\rm det}'
\left[-\Box_E+U''(\phi_b)\right
]}{{\rm det}\left[-\Box_E+U''(\phi_f)\right]}\right]\right\}\nonumber\\
&=&\exp\left\{-\frac{1}{2}\ln\left[\frac{\prod_i\,'E_b^2(i)}
{\prod_{j}E_f^2(j)}\right]\right\}\nonumber\\
&=&\exp\left\{-\frac{1}{2}\left[\sum_{i}\,'
\ln |E_b^2(i)|-\sum_{j}
\ln E_f^2(j)\right]\right\} \;.
\end{eqnarray}

Since the bounce can be approximated by a constant field configuration for 
$r<(R-\Delta R)\approx R$,
we can write for $K$, in the thin wall approximation, the expression:

\begin{equation}
K=\exp\left\{-\frac{1}{2}\left[\frac{4}{3}\pi R^3\int \frac{d^3p}{(2\pi)^3}
\ln\left[\frac{E_t^2(p)}{E_f^2(p)}\right]+\left(\sum_{i}\,'
\ln |E_{\rm wall}^2(i)|-\sum_{j}\ln E_f^2(j)\right)\right]
\right\}
\;,
\label{pion}
\end{equation}

\noindent
where $E_{t(f)}^2(p)=p^2+U''(\phi_{t(f)})$. 
The integral  in (\ref{pion}) can then be identified as the one loop 
correction 
to the classical potential, while the remaining terms represent the 
quantum
corrections due to fluctuations around the bounce wall \cite{guenther}. 
Then, by using Eqs. (\ref{acao}) and (\ref{pion}) in 
Eq. (\ref{decayrate}) and following \cite{coleman,callan}, we obtain

\begin{equation} 
\frac{\Gamma}{V_3} \simeq 2\left(
\frac{\Delta S_E}{2\pi}\right)^{3/2}
\exp\left[\frac{4}{3}\pi R^3\Delta U_{eff}-4\pi 
R^2\left({\cal S}_0+{\cal S}_1\right)\right] \;,
\label{taxa2} 
\end{equation}

\noindent
where ${\cal S}_1$, is the term giving the 1-loop quantum corrections to 
fluctuations 
around the bounce wall,

\begin{equation} 
{\cal S}_1=\frac{1}{4\pi R^2}\left(\sum_{i}\,'\ln \left|E_{\rm wall}^2(i) 
\right|-
\sum_{j}\ln E_f^2(j)\right) \;,
\label{libra}
\end{equation}

\noindent
where $E_{\rm wall}^2(i)$ are the eigenvalues of 
$-\Box_E+U''(\phi_{\rm wall}(r-R)$. In (\ref{taxa2}) we have also that 
$\Delta  U_{\rm eff}=
U_{\rm eff}(\phi_t)-
U_{\rm eff}(\phi_f)$,
where $U_{\rm eff}$
is the one loop effective potential, given by \cite{swanson}

\begin{equation}
U_{\rm eff}(\phi)=U(\phi)+\frac{1}{2}\int \frac{d^3 p}{(2\pi)^3} 
\ln\left[\frac{p^2+U''(\phi)}{p^2+m^2}\right] \;.
\label{poteff}
\end{equation}

\noindent
The ultraviolet divergence 
in (\ref{poteff}) can be handled in the usual way.
Integrating over $p_0$ and
by using an ultraviolet
cut-off, $\Lambda$, for the space-momentum, we obtain
 
\begin{equation}
U_{\rm eff}(\phi)=U(\phi)-\frac{1}{12}
\left[\left({\bf{p}}^2+m^2\right)^{3/2}\Big|_{0}^{\Lambda}-\left({\bf{p}}^2+
m^2+\frac{\lambda}{2}\phi^2+\frac{\sigma}{4!}\phi^4\right)^{3/2}
\Big|_{0}^{\Lambda}\right] \;.
\end{equation}

\noindent
Using that $(1+x)^{3/2}=1+3/2x+3/8x^2+ \ldots$, 
we obtain for $U_{\rm eff}$ the expression

\begin{eqnarray}
U_{\rm eff}(\phi)&=&U(\phi)-\frac{1}{12}\left[\left(m^2+
\frac{\lambda}{2}\phi^2+
\frac{\sigma}{4!}\phi^4\right)^{3/2}-m^3-\frac{3}{2}Y(\Lambda)
\left(\frac{\lambda}{2}\phi^2+\frac{\sigma}{4!}\phi^4\right) \right]
\nonumber \\
& &+ {\cal O}\left(1/\Lambda\right) \;,
\label{pujada}
\end{eqnarray}

\noindent
where $Y(\Lambda)=\Lambda^2+m_r^2$. 
The divergent terms in (\ref{pujada}) are proportional 
to 
$\phi^2$ and $\phi^4$ but not to $\phi^6$. Then, only the mass $m$ and 
$\lambda$ 
need to be renormalized ($\sigma_r=\sigma$). {}From the usual definition of 
renormalized mass $m_r$ 
and coupling constant $\lambda_r$,

\begin{equation}
m_{r}^2=\frac{d^2U_{\rm eff}(\phi)}{d\phi^2} \Big|_{\phi=0}
 \label{masar}
\end{equation}

\noindent
and 

\begin{equation}
\lambda_{r}=\frac{d^4U_{\rm eff}(\phi)}{d\phi^4} \Big|_{\phi=0}
\label{lambdar}
\end{equation}

\noindent
and writing the unrenormalized parameters in terms of renormalized ones,
we obtain
for the renormalized one loop effective potential, the following expression:

\begin{equation}
U_{\rm eff}(\phi)=U_r (\phi)-\frac{1}{12}\left[U_r ''(\phi)^{3/2}-m_r^3-
\frac{3}{4}m_r \lambda_r \phi_r^2-\frac{3\lambda_r^2}{32m_r}\phi^4-
\frac{3m_r\sigma_r}{16}\phi^4\right]\;,
\end{equation}

\noindent
where $U_r$ means the tree level potential expressed in terms of the 
renormalized
quantities. {}For convenience, from now on we drop the $r$ subscript from
the expressions and it is to be understood that the parameters  
$m,\lambda$ and $\sigma$ are
the renormalized ones, instead of the bare ones. Then,
$\Delta U_{\rm eff}$ can be written as (since $\phi_f=0$)

\begin{equation}
\Delta U_{\rm eff}=U(\phi_t)-\frac{1}{12}\left[U''(\phi_t)^{3/2}-
\frac{3}{4}m\lambda\phi_t^2-\frac{3\lambda^2}{32m}\phi_t^4-
\frac{3m\sigma}{16}\phi_t^4\right]
\end{equation}

We now turn to the problem of evaluating the eigenvalues 
$E_{\rm wall}^2(i)$ of $-\Box_E+U''[\phi_{\rm wall}(r-R)]$, 
which appears in
(\ref{pion}). This is not an easy task. In fact, only in a very few 
examples this has known analytical solutions, as, for example, for the 
kink solution in the $(\lambda \phi^4)_{D=2}$ 
model \cite{rajaraman}. Unfortunately, for the model studied here, we can 
not find analytical solutions for these differential operators.
However, we can perform an approximate analysis, and, in particular we can find   
explicitly the negatives and zero modes for the differential operator for the 
bounce wall field configuration. 
By making use of the 
spherical symmetry of the bounce solution, we can express the eigenvalue 
equation:

\[
[-\Box_E+U''(\phi_{\rm wall}(r-R))]\Psi_i(r,\theta,\varphi)=
E_{\rm wall}^2(i)\Psi_i(r,\theta,\varphi)
\]

\noindent
in the form

\begin{equation}
\left[-\frac{d^2}{dr^2}-\frac{2}{r}\frac{d}{dr}+\frac{l(l+1)}{r^2}+
m^2+\frac{\lambda}{2}\phi_{\rm wall}^2(r-R)+\frac{\sigma}{4!}
\phi_{\rm wall}^4
(r-R)\right]\psi_i(r)=E_{\rm wall}^2(i)\psi_i(r) \;,
\label{auto1}
\end{equation}

\noindent
where $l=0,1,2,\ldots$. Making $\psi_i(r)=\chi_i(r)/r$ and $z=r-R$, 
we obtain

\begin{equation}
\left[-\frac{d^2}{dz^2}+\frac{l(l+1)}{(z+R)^2}+m^2+\frac{\lambda}{2}
\phi_{\rm wall}^2(z)+\frac{\sigma}{4!}\phi_{\rm wall}^4(z)\right]
\chi_{n,l}(z)=
E_{\rm wall}^2 (n,l)\chi_{n,l}(z)\;.
\label{auto2}
\end{equation}

\noindent
Since $\Delta R\ll R$, we can take
$l(l+1)/(z+R)^2\approx l(l+1)/R^2$ and then

\begin{equation}
\left[-\frac{d^2}{dz^2}+\frac{\lambda}{2}\phi_{\rm wall}^2(z)+
\frac{\sigma}{4!}\phi_{\rm wall}^4(z)\right]\chi_n(z)=\eta_n^2 \chi_n(z)\;,
\label{auto3}
\end{equation}

\noindent
where $\eta_n$ is obtained from

\begin{equation}
E_{\rm wall}^2(n,l)=\eta_n^2 +m^2+\frac{l(l+1)}{R^2}\;.
\label{auto4}
\end{equation}

We know that the $-\Box_E+U''(\phi_b(r))$ operator has three zero 
eigenvalues 
coming from the bounce translational invariance. 
Then, for $l=1$ and to the lowest value of $\eta_n$ (which can be chosen as $\eta_1$), 
we will have $E_{\rm wall}^2 (1,1)=0$, with multiplicity three, as expected,  and  
$\eta_1^2=-m^2-2/R^2$. 
The lowest eigenvalue $E_{\rm wall}^2$ (the negative eigenvalue) will be 
$E_{\rm wall}^2(1,0)=-2/R^2$, with multiplicity one, just what one
would expect for the metastable state, the existence of only one negative
eigenvalue \cite{guenther}. 
To evaluate the other eigenvalues, we make the following change of variable 
$w=\sqrt{\frac{\sigma}{6!}}\phi_0^2 \, z$ and use (\ref{wall}) in 
(\ref{auto4}).  
We then get

\begin{equation}
\left[-\frac{d^2}{dw^2}-\frac{24}{1+e^{\sqrt{8}w}}+\frac{30}{\left(1+
e^{\sqrt{8}w}\right)^2}\right]\chi_n(w)=\nu_n^2 \chi_n(w) \;,
\label{auto5}
\end{equation}

\noindent
where $\nu_n^2=\eta_n^2 \, 6!/(\sigma\phi_0^4)$. We can then express
the eigenvalues 
of $-\Box_E+U''(\phi_{\rm wall}(r-R))$ as

\begin{equation}
E_{\rm wall}^2(n,l)=\phi_0^4\frac{\sigma}{6!}\,\nu_n^2+
m^2+\frac{l(l+1)}{R^2} \;.
\label{auto6}
\end{equation}

We were not able to find any analytical solution for Eq. (\ref{auto5}),
which it is even harder to solve, due to the boundary conditions. We are
currently working on the numerical solution for the eigenvalues, 
whose results
will be reported elsewhere.

\section{The Nucleation Rate at Finite Temperature}

At finite temperature, the bounce solution is a static solution of the field
equation of motion, $\delta S_E/\delta \phi |_{\phi = 
\phi_B (\rho)}=0$, 
where $\phi_B(\rho)$ is given as in (\ref{bounce}), with 
bubble radius $\bar{\rho}$ and thickness $\Delta \rho \ll \bar{\rho}$.
$\phi_{\rm wall} =
\phi_{\rm wall} (\rho)$
is still expressed as in (\ref{wall}).
To calculate the determinantal prefactor in
(\ref{reconcha}), we consider the eigenvalue equations for the
differential operators:

\begin{equation}
\left[-\Box_{E} +U''(\phi_B)\right]\psi_B(i)=\mu_B^2\psi_B(i)
\label{papi2}
\end{equation}

\noindent
and

\begin{equation}
\left[-\Box_{E} +U''(\phi_f)\right]\psi_f(i)=\mu_f^2\psi_f(i) \;,
\label{mami2}
\end{equation}

\noindent
where, in momentum space, $ \mu^2=\omega_n^2+E^2 $, where
$\omega_n=\frac{2\pi n}{\beta}$ are the Matsubara frequencies 
($n=0,\pm 1,\pm 2,\ldots$). Using (\ref{papi2}) and (\ref{mami2}) 
in (\ref{reconcha}), we obtain for $K$ the expression:

\begin{eqnarray}
K_{\beta}&=&\left[\frac{{\rm det}\left[-\Box_E+U''(\phi_B)\right
]_{\beta}}{{\rm det}\left[-\Box_E+U''(\phi_f)\right]_{\beta}}\right]^{-1/2}
\nonumber\\
&=&\exp\left\{-\frac{1}{2}\ln\left[\frac{{\rm det}\left[-\Box_E+U''(\phi_B)
\right
]_{\beta}}{{\rm det}\left[-\Box_E+U''(\phi_f)\right]_{\beta}}\right]\right\}
\nonumber\\
&=&\exp\left\{-\frac{1}{2}\ln\left[\frac{\prod_{n=-\infty}^{+\infty}
\prod_{i}\left[\omega_n^2+E_B^2(i)\right]}
{\prod_{n=-\infty}^{+\infty}\prod_{j}\left[\omega_n^2+
E_f^2(j)\right]}\right]\right\}\nonumber\\
&=&\exp\left\{-\frac{1}{2}\ln\left[\frac{\prod_{n=-\infty}^{+\infty}
\left(\omega_n^2+E_-^2\right)\left(\omega_n^2+E_0^2\right)^2
\prod_{i}^{'}\left[\omega_n^2+E_B^2(i)\right]}
{\prod_{n=-\infty}^{+\infty}\prod_{j}\left[\omega_n^2+
E_f^2(j)\right]}\right]\right\}\;,
\label{zeros}
\end{eqnarray}

\noindent 
where we have separated the negative and zero eigenvalues in
the numerator of Eq. (\ref{zeros}), with the prime meaning that the single
negative eigenvalue, $E_-^2$, and the two zero eigenvalues, $E_{0}^2$ (related to the now
two-dimensional space), were
excluded from the product. The term for $n=0$ in $
\left(\omega_n^2+E_0^2\right)$ can be handled using collective
coordinates method, as in \cite{callan,rajaraman}, resulting in the factor $
V_2\left[\frac{{\Delta}E}{2\pi T}\right]$, where $V_2$ is 
the ``volume''  of the
two-space. Separating the $n=0$ modes from both the numerator and
denominator of (\ref{zeros}) and using the identity 

\begin{equation}
\prod_{n=1}^{n=+\infty}\left(1+\frac{a^2}{n^2}\right)=\frac{\sinh(\pi a)}
{\pi a}\;,
\label{recurso}
\end{equation}

\noindent
we get

\begin{equation}
K_{\beta}=V_2 \left[\frac{{\Delta}E}{2\pi T}\right]
\exp\left\{-\ln(E_-^2)^{\frac{1}{2}}-\ln\left[\frac{\sin(\frac{\beta}{2}
|E_-|)}{\frac{\beta}{2}|E_-|}\right]+a+b\right\}\;,
\label{abc}
\end{equation}

\noindent
where

\begin{equation}
a=\left(-3+\sum_{j}-\sum_{i}~'\right)
\ln\prod_{n=1}^{+\infty}\omega_{n}^2+\left(\sum_{i}~'
-\sum_{j}\right)\ln\beta 
\label{antro}
\end{equation}

\noindent
and

\begin{equation}
b=\sum_{j}\left[\frac{\beta}{2}E_f(j)+
\ln\left(1-e^{-\beta E_f(j)}\right)\right]-
\sum_{i}~'\left[\frac{\beta}{2}E_B(i)+
\ln\left(1-e^{-\beta E_B(i)}\right)\right]\;.
\label{baca}
\end{equation}

\noindent
Since $\sum_{i}~'$ has three eigenvalues less than 
$\sum_{j}$ and $(E_-^2)^{1/2}=i |E_-|$, we obtain

\begin{equation}
K_{\beta}=-i\, V_2 \left[\frac{{\Delta}E}{2{\pi} T}
\right]\left[2\beta^2\sin\left(\frac{\beta}{2}|E_-|\right)\right]^{-1}e^b 
\;. 
\label{pana}
\end{equation}

\noindent
Using the above results in (\ref{reconcha}), 
we then obtain
for the nucleation rate, per unit volume, the expression:

\begin{equation}
\frac{\Gamma_\beta}{V_2} = 2QT^3\exp\left[-\frac{\Delta
F(T)}{T}\right] \;,
\label{graff}
\end{equation}

\noindent
with

\begin{equation}
Q=\left(\frac{{\Delta}E}{2{\pi}T}\right)
\frac{\frac{|E_{-}|}{2T}}{\pi\sin(\frac{|E_{-}|}{2T})}
\end{equation}

\noindent
and

\begin{equation}
{\Delta} F(T)={\Delta}
E-\sum_{j}\left[\frac{1}{2}E_f(j)+\frac{1}{\beta}
\ln\left(1-e^{-\beta E_f(j)}\right)\right]+\sum_{i}~'
\left[\frac{1}{2}E_B(i)
+\frac{1}{\beta}\ln\left(1-e^{-\beta E_B(i)}\right)\right]\;.
\end{equation}

In the thin-wall approximation (in an analogous way as was done in Sec. III) we can write

\begin{equation}
{\Delta}F(T) \simeq -\pi {\bar \rho}^2 \, \Delta U_{\rm eff}(T)+
2\pi ({\cal S}_0+{\cal S}_{\beta})  \bar{\rho} \;,
\label{bomba}
\end{equation}

\noindent
where $\Delta U_{\rm eff} (T)$ is given by

\begin{eqnarray}
\Delta U_{\rm eff}(T) &=&\epsilon_0+\int\frac{d^2{\bf{p}}}
{(2\pi)^2}\left[\frac{1}{2}
\sqrt{{\bf{p}}^2+U''(\phi_f)}-\frac{1}{2}\sqrt{{\bf{p}}^2+
U''(\phi_t)}~\right]
\nonumber\\
& &+T\int\frac{d^2{\bf{p}}}{(2\pi)^2}
\ln\left[1-e^{-\beta\sqrt{{\bf{p}}^2+U''(\phi_f)}}~\right]-
T\int\frac{d^2{\bf{p}}}{(2\pi)^2}
\ln\left[1-e^{-\beta\sqrt{{\bf{p}}^2+U''(\phi_t)}}~\right]\nonumber\\
\label{efectivo}
\end{eqnarray}

\noindent
and ${\cal S}_\beta$ is the 1-loop finite temperature correction to the 
bubble surface energy density, given by

\begin{equation}
{\cal S}_{\beta}=\frac{T}{2\pi \bar{\rho}}\left\{\sum_{i}~'
\left[\frac{\beta}{2}E_{\rm wall}(i)+
\ln\left(1-e^{-\beta E_{\rm wall}(i)}\right)\right]-
\sum_{j}\left[\frac{\beta}{2}E_f(j)+
\ln\left(1-e^{-\beta E_f(j)}\right)\right]\right\} \;,
\label{nulo}
\end{equation}

\noindent 
where $E_{\rm wall}^2(i)$ are the eigenvalues of
$-{\bf \nabla}^2 +U''(\phi_{wall}(\rho-\bar{\rho}))$. In (\ref{efectivo}), the
first integral is divergent but it can be handled just in the same way
as in the previous section, by the
introduction of the appropriated counterterms of renormalization.
The remaining integrals in (\ref{efectivo}) are all finite and they
can be reduced to integrals of the type

\begin{equation}
I(t)=\int_0^{\infty}dx
\left[x\ln\left(1-e^{\sqrt{x^2+t^2}}~\right)\right]
\label{Int}
\end{equation}

\noindent
and it is evaluated in Appendix B. The result is:

\begin{equation}
I(t)=I(0)+\frac{t^3}{6}+\frac{t^2}{4}-\frac{t^2}{2}\ln{t}-
\frac{1}{2}\sum_{n=1}^{+\infty}\frac{(-1)^{n+1}\zeta(2n)}
{n(n+1)(2\pi)^{2n}}\left(t^2\right)^{n+1}\;,
\label{intI}
\end{equation}

\noindent
where $\zeta(n)$ is the Riemman Zeta function. Using (\ref{intI}) in
(\ref{efectivo}), we obtain for (\ref{bomba}) the following expression

\begin{equation}
{\Delta}F(T)=-\pi \bar{\rho}^2 \left[\epsilon_0+c(\phi_f)-c(\phi_t)\right]+
2\pi \bar{\rho} \left[{\cal S}_0+{\cal S}_{\beta}\right] \;,
\label{dina}
\end{equation}

\noindent
where (using $t=\beta U''(\phi)^{1/2}$)

\begin{equation}
c(\phi)=\frac{1}{2\pi \beta^3}I\left(t\right)
\simeq\frac{1}
{2\pi \beta^3}\left(I(0)+\frac{t^3}{6}+\frac{t^2}{4}-\frac{t^2}{2}\ln{t}
-\frac{\zeta(2)t^4}{16\pi^2}\right)\;.
\end{equation}

\noindent
The critical radius for bubble nucleation, $\rho_c$, is obtained by minimizing (\ref{dina}),
$\delta \Delta F(T) / \delta \bar{\rho} |_{\bar{\rho}= \rho_c} = 0$.

{}For the eigenvalues $E_{\rm wall}^2(i)$ of $-{\bf \nabla}^2
+U''(\phi_{\rm wall}(\rho-\bar{\rho}))$, using the now cylindrical symmetry 
of $\phi_{\rm wall}(\rho)$, the eigenvalue equation

\[
[-{\bf \nabla}^2 +U''(\phi_{\rm wall}(\rho-\bar{\rho}))]\Psi_i(\rho,\varphi)=
E_{\rm wall}^2(i)\Psi_i(\rho,\varphi)
\]

\noindent
can  be written as

\begin{equation}
\left[-\frac{d^2}{d\rho^2}-\frac{1}{\rho}\frac{d}{\rho}+
\frac{s^2}{\rho^2}+m^2+
\frac{\lambda}{2}\phi_{\rm wall}^2(\rho-\bar{\rho})+\frac{\sigma}{4!}
\phi_{\rm wall}^4(\rho-\bar{\rho})\right]\psi_{n,s}(\rho)=E_{\rm wall}^2
(n,s)\psi_{n,s}(\rho)\;,
\label{peche}
\end{equation}

\noindent
where $s=0,\pm 1, \pm 2, \ldots$. Taking
$\psi(\rho)=\chi(\rho)/\rho^{1/2}$, 
we obtain

\begin{equation}
\left[-\frac{d^2}{d\rho^2}+\frac{4s^2-1}{4\rho^2}+m^2+
\frac{\lambda}{2}\phi_{\rm wall}^2(\rho-\bar{\rho})+\frac{\sigma}{4!}
\phi_{\rm wall}^4(\rho-\bar{\rho})\right]\chi_{n,s}(\rho)=E_{\rm wall}^2
(n,s)\chi_{n,s}(\rho)\;.
\label{peche2}
\end{equation}

As in the previous section, we make $z=(\rho-\bar{\rho})$ and because 
$\Delta \rho \ll \bar{\rho}$, then 

\begin{equation}
\left[-\frac{d^2}{dz^2}+\frac{\lambda}{2}\phi_{\rm wall}^2(z)+
\frac{\sigma}{4!}\phi_{\rm wall}^4(z)\right]\chi_n(z)=
\eta_n^2\chi_n(\rho)\;,
\label{peche3}
\end{equation}

\noindent
where $\eta_n^2$ is now obtained from

\begin{equation}
E_{wall}^2(n,s)=\eta_n^2+m^2+\frac{4s^2-1}{4\bar{\rho}^2}\;.
\label{val5}
\end{equation}

\noindent
Analogously to the zero temperature case, the differential operator
$-{\bf \nabla}^2 +U''(\phi_B)$ has now two zero
eigenvalues (related to the translational modes in the 
two-dimensional space). The multiplicity of (\ref{val5}) is two for 
all $s\neq 0$
and for $s=0$ the multiplicity is one. Then for $s=1$ and lower
$\eta_n$ (we choose $\eta_1$) we will have $E_{\rm wall}^2(1,1)=0$
and then $\eta_1^2=-[m^2+3/(4\bar{\rho}^2)]$. {}For
the negative eigenvalue we obtain
$E_-^2=E_{\rm wall}^2(1,0)=-1/\bar{\rho}^2$. As in the previous
section, by taking $w=\sqrt{\sigma/6!} \, \phi_0^2 \, z$, we obtain for the
eigenvalues, 

\begin{equation}
E_{wall}^2(n,s)=\phi_0^4\frac{\sigma}{6!}\, \nu_n^2 +m^2+\frac{4s^2-1}
{4\bar{\rho}^2}\;,
\label{val6}
\end{equation}

\noindent
where $\nu_n$ is the same as in the previous section.

\section{Conclusions}

In this paper we have studied the evaluation of the vacuum decay rates, 
at both 
zero temperature and at finite temperature, for the $\sigma \phi^6$ model
in $D=3$, when the parameters of the model satisfies the conditions given
in (\ref{param}). Our main results were the determination of the
expression for the bounce solution, Eqs. (\ref{bounce}) and (\ref{wall}),
and also, despite of the difficulties for finding the solutions for the 
bounce's wall
eigenvalue problem, by taking consistent considerations for the field 
equations we have at hand (like the thin-wall approximation), we were
able to perform a detailed analysis of the bounce negative and zero
eigenvalues. We have also given a set of eigenvalue equations, which can
be useful in a more detailed analysis of this problem using, {\it e.g.} 
numerical methods. 
In \cite{amaral} it was analyzed the phase structure for the $\sigma \phi^6$ 
model in $D=2$ in the lattice and it was also studied the possible 
production  of topological and nontopological excitations in the model.
By remembering that at high temperatures our model in $D=3$ resembles the 
$D=2$ model at zero temperature,
it will be interesting to apply the method and results we have 
obtained here to the problem studied in \cite{amaral} for the regime of 
a first order phase transition.

\vspace{0.5cm}

\begin{center}
{\large \bf Acknowledgements}
\end{center}

ROR and NFS are partially supported by Conselho Nacional de
Desenvolvimento Cient\'{\i}fico e Tecnol\'ogico - CNPq (Brazil). GHF 
is supported by a grant from CAPES. 
ROR would like also to thank ICTP-Trieste, for
the kind hospitality, when during his stay, this work was completed.

\appendix

\section*{Appendix}

%\newcommand{\alpheqn}%
 %{\setcounter{saveeqn}{\value{equation}}%
%\stepcounter{saveeqn}\setcounter{equation}{0}%
%\renewcommand{\theequation}%
 %{\mbox{\arabic{section}.\arabic{saveeqn}\alph{equation}}}}
%\newcommand{\reseteqn}{\setcounter{equation}{\value{saveeqn}}%
\renewcommand{\theequation}{\mbox{A.\arabic{equation}}}

We evaluate here the integral (\ref{Int}),

\begin{equation}
I(t)=\int_0^{\infty}dx
\left[x\ln\left(1-e^{\sqrt{x^2+t^2}}~\right)\right]\;.
\end{equation}

\noindent
Using that

\begin{equation}
\frac{\partial I(t)}{\partial
(t^2)}=\int_0^{\infty}dx\left[x\frac{\partial }{\partial (t^2)}
\ln\left(1-e^{\sqrt{x^2+t^2}}~~\right)\right]=
\int_0^{\infty}dx\left[x\frac{\partial }{\partial (x^2)}
\ln\left(1-e^{\sqrt{x^2+t^2}}~~\right)\right]
\end{equation}

\noindent
and integrating by parts, we obtain that

\begin{equation}
\frac{d I(t)}{d(t^2)}=\frac{1}{2}\ln\left(1-e^{\sqrt{x^2+t^2}}~~\right)
\Big|_0^{\infty}=
-\frac{1}{2}\ln\left(1-e^{-t}\right)\;.
\nonumber
\end{equation}

\noindent
We can then write $I(t)$ in the form:

\begin{equation}
I(t)=I(0)-\int_0^tdt\left[ t\ln\left(1-e^{-t}\right)\right]=
I(0)-\int_0^tdt \left[t\ln\sinh
\left(\frac{t}{2}\right)+t\ln2-\frac{t^2}{2}\right]\nonumber
\end{equation}

\noindent
If one uses (\ref{recurso}) and expanding the logarithm in the above
equation, we are able
to perform the integral and we obtain the result shown in Sec. IV, 
Eq. (\ref{intI}).

\newpage

\begin{center}
{\large \bf Figure Captions}
\end{center}

\vspace{1.cm}

{\bf Figure 1:}The potential $U(\phi)$ for parameters $m,\lambda,\sigma$ 
satisfying Eq. (1.1).

{\bf Figure 2:}The inverted potential $-U(\phi)$.

{\bf Figure 3:}The bubble field configuration $\phi_{\rm wall}(r)$.

\begin{figure}[tb]
\centerline{\epsfysize=8in\epsffile{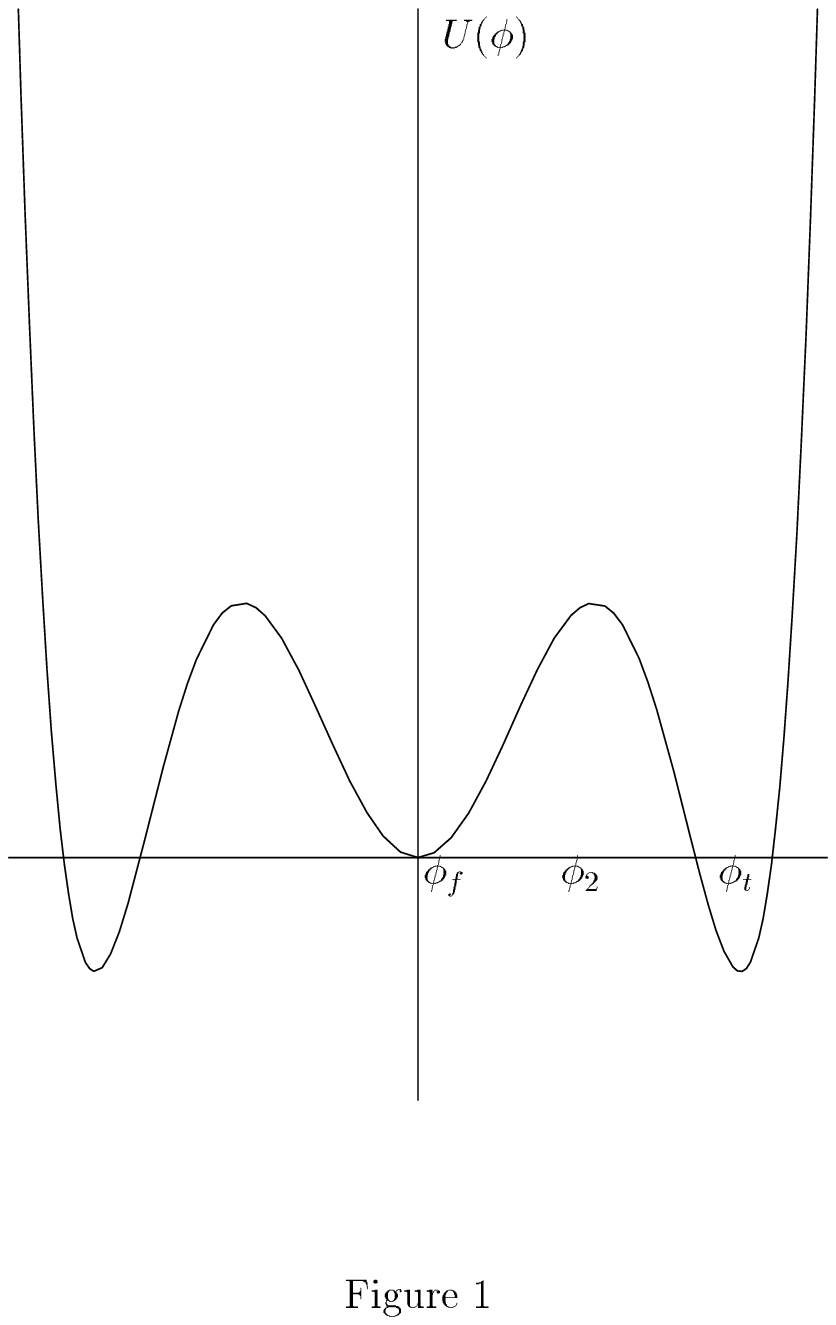}}
%\caption[region]
%{\small\sf{The potential $U(\phi)$ for $m=,\lambda=,\sigma=$.}}
\end{figure}

\begin{figure}[tb]
\centerline{\epsfysize=8in\epsffile{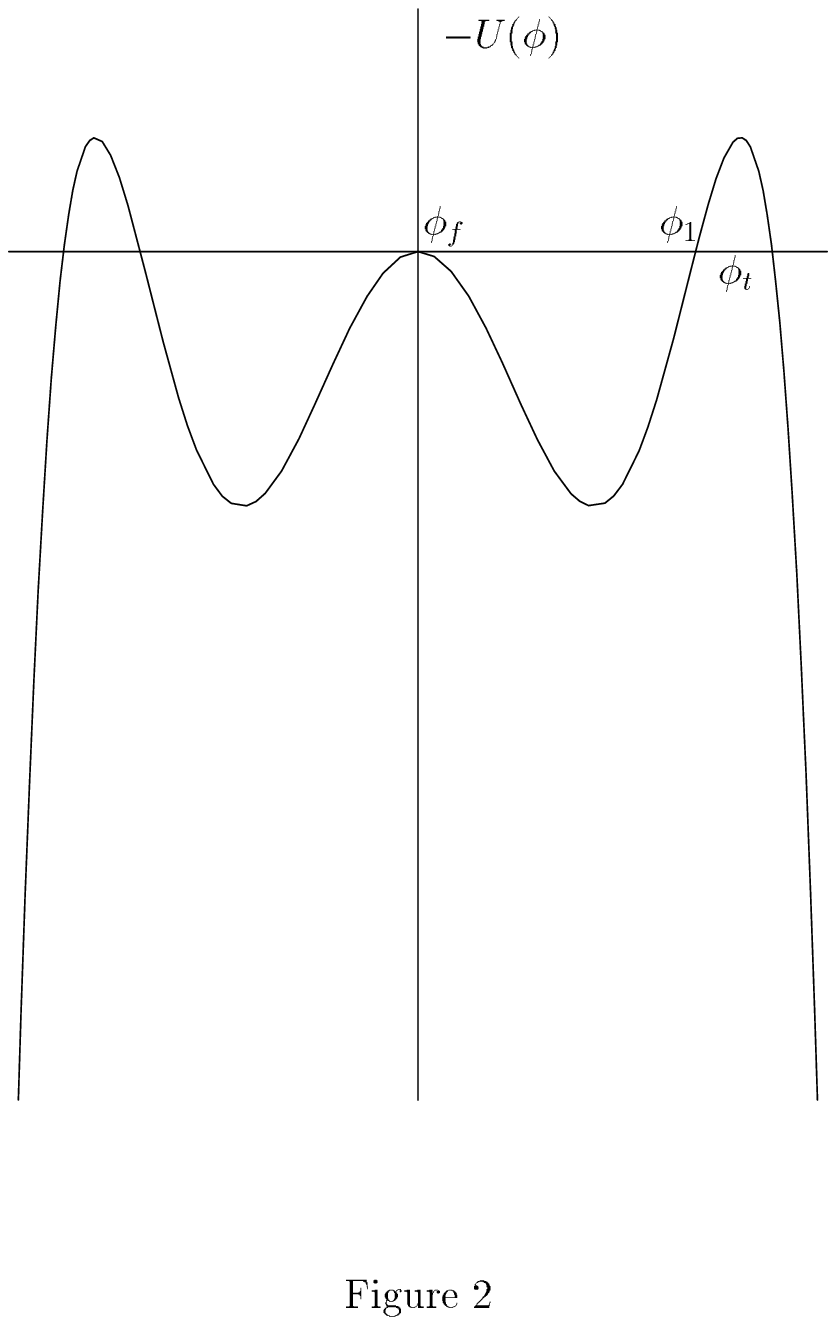}}
%\caption[region]
%{\small\sf{The inverted potential $-U(\phi)$.}}
\end{figure}

\begin{figure}[tb]
\centerline{\epsfysize=8in\epsffile{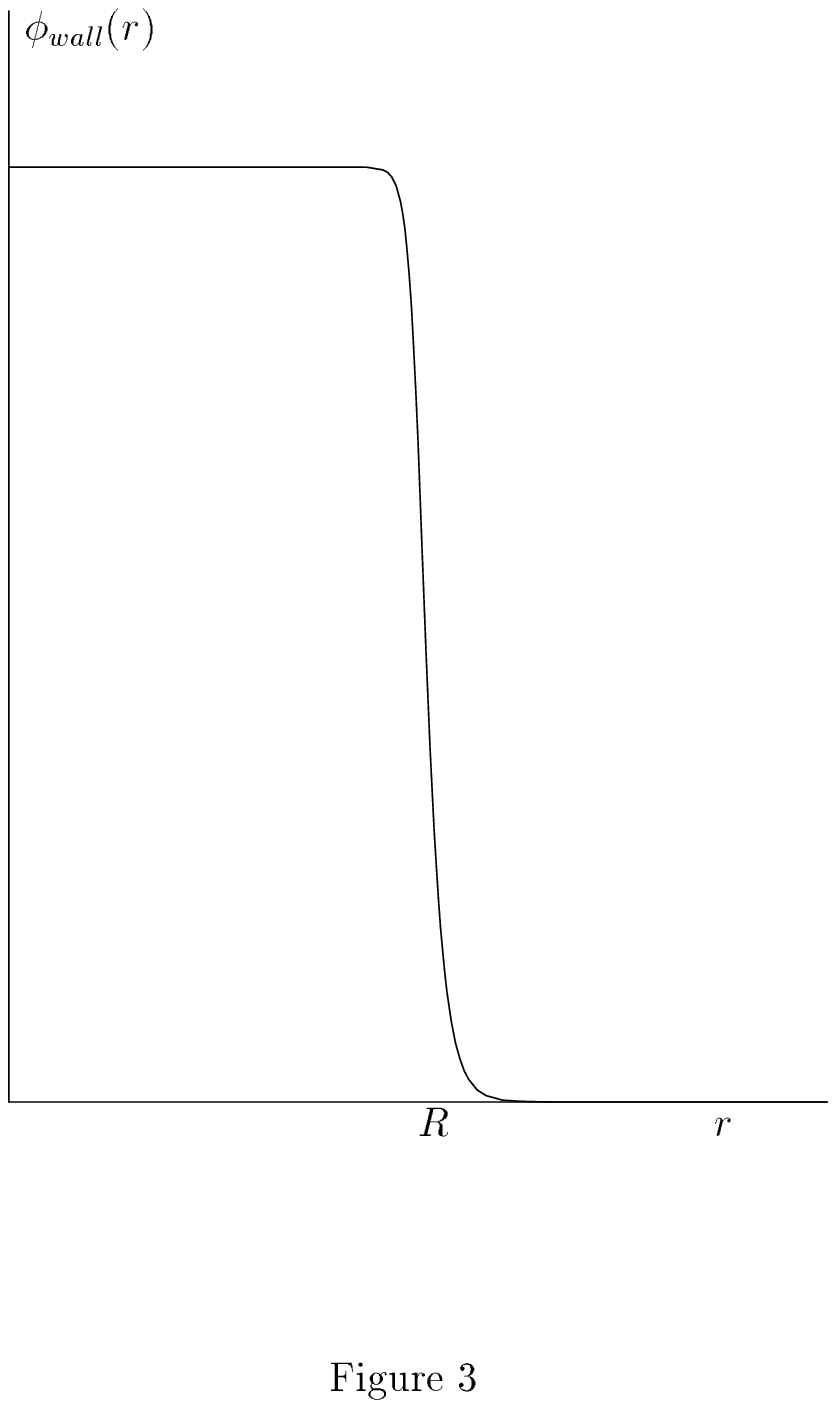}}
%\caption[region]
%{\small\sf{The bubble field configuration $\phi_{\rm wall}(r)$.}}
\end{figure}

\end{document}